\documentclass[a4paper]{article}
\usepackage{iftex}
\ifLuaTeX\else\pdfoutput=1\fi
\usepackage{hyperref}
\newcommand{\R}{\ensuremath{\mathbb{R}}}
\usepackage{amsmath,amssymb}
\usepackage{stmaryrd}
\usepackage{graphicx}
\usepackage{subfiles}
\usepackage{braket}
\usepackage{cleveref}
\usepackage[prologue,gray]{xcolor}
\usepackage{iftex}
\ifLuaTeX
\usepackage{fontspec}
\else
\usepackage{inconsolata}
\fi
\usepackage{tikz}
\usetikzlibrary{matrix,decorations,arrows,calc,trees}
\usepackage{tcolorbox}
\tcbuselibrary{listings,xparse}
\usepackage[frozencache]{minted}
\setminted{style=borland}

\definecolor{ltblue}{rgb}{0,0.4,0.4}
\definecolor{dkblue}{rgb}{0,0.1,0.6}
\definecolor{dkgreen}{rgb}{0,0.35,0}
\definecolor{dkviolet}{rgb}{0.3,0,0.5}
\definecolor{dkred}{rgb}{0.5,0,0}
\definecolor{comment}{HTML}{444444}
\definecolor{keywd}{HTML}{8D00ED}
\definecolor{types}{HTML}{1F7B2F}
\definecolor{str}{HTML}{4070a0}
\definecolor{code-background}{gray}{0.8}
\definecolor{pragma}{HTML}{372A78}
\definecolor{num}{HTML}{40a070}
\definecolor{symb}{HTML}{000000}

\let\textt=\texttt
\newminted[code]{haskell}{%
mathescape,linenos,numbersep=5pt,frame=lines,framesep=2mm,%
}
\newminted[repl]{haskell}{%
mathescape,numbersep=5pt,frame=lines,framesep=2mm,%
}
\VerbatimFootnotes
\DeclareTotalTCBox{\hask}{v}{%
tcbox raise base,box align=base,verbatim,colback=lightgray,colframe=gray%
}{\mintinline[fontsize=\small]{haskell}{#1}}

\usepackage{pgfplots}
\usepackage[backend=biber,bibstyle=numeric]{biblatex}
\Crefformat{enumi}{(#2#1#3)}
\pgfplotsset{small,ylabel style={yshift=-12pt},%
  legend style={%
    outer sep=7pt,inner sep=5pt,%
    /tikz/every even column/.append style={column sep=10pt},%
  },
}

\title{A Succinct Multivariate Lazy Multivariate Tower AD for Weil Algebra Computation}
\newcommand{\Rseries}{\R\llbracket\boldsymbol{X}\rrbracket}
\date{}

\usepackage{geometry}
\newenvironment{acknowledgements}{\section*{Acknowledgements}}{}

\author{Hiromi Ishii}

\begin{document}
\maketitle

\begin{abstract}
  We propose a functional implementation of \emph{Multivariate Tower Automatic Differentiation}.
  Our implementation is intended to be used in implementing $C^\infty$-structure computation of an arbitrary Weil algebra, which we discussed in \cite{Ishii:2021vw}.
\end{abstract}

\frenchspacing

\section{Introduction}
\label{sec:intro}
\emph{Automatic Differentiation} (AD) is known as a powerful technique to compute differential coefficients of a given (piecewise) smooth function efficiently and accurately.
In the upcoming paper~\cite{Ishii:2021vw}, the author proposed to use \emph{$C^\infty$}-rings and Weil algebras to provide a modular and exprresive framework for forward-mode automatic difrerentiation.
There, compute the $C^\infty$-structure of an arbitrary Weil algebra as a quotient of that of the formal power series ring $\Rseries$.
The $C^\infty$-structure of $\Rseries$ was then computed via \emph{multivariate tower AD}.
It can be implemented in various ways, such as Lazy Multivariate Tower AD~\cite{Pearlmutter:2007aa}, or nested Sparse Tower AD~\cite[{module \texttt{Numeric.AD.Rank1.Sparse}}]{Kmett:2010aa}.

Theoretically, such existing methods can be used to compute the $C^\infty$-structure of $\Rseries$.
However, these methods are somewhat complex and not optimised for our purpose.
In this paper, we will propose another implementation of Lazy Multivariate Tower-Mode AD using tree representation and exploiting smoothness to save memory consumption.
Our method can be seen as aforementioned existing implementations~\cite{Pearlmutter:2007aa,Kmett:2010aa}.

\section{Implementation}
\label{sec:impl}
As an implementation language, we adopt a Haskell~\cite{haskell.org:2021tt}, a purely functional lazy programming language.
It has several virtues useful for our purpose:
\begin{enumerate}
\item It supports higher-order functions natively.
\item It is a lazy language, enabling us to treat \emph{infinite} structures.
\item The type-class mechanism in Haskell allows us to use function overloading in handy.
\item Its type system allows us to implement complex inductive types safely.
\end{enumerate}
For a general discussion on the advantages of using Haskell in computer algebra, we refer readers to Ishii~\cite{ISHII:2018ek}.

We follow the standard pattern in implementing ADs in Haskell: we use the function overloading to implement operations on types corresponding ADs.
This strategy is taken, for example, in Karczmarczuk~\cite{Karczmarczuk:2001ww}, Elliott~\cite{Elliott2009-beautiful-differentiation} and implemented in \texttt{ad} package~\cite{Kmett:2010aa}.
In Haskell, the \hask{Floating} type-class gives an abstraction over floating-point numbers that admit elementary functions, as excerpted in \Cref{lst:cls-floating}.
\begin{listing}[tbp]
\begin{code}
class Fractional a => Floating a where
  pi :: a
  exp :: a -> a
  log :: a -> a
  sin :: a -> a
  asin :: a -> a
  ...
\end{code}
\caption{The \texttt{Floating} class\label{lst:cls-floating}}
\end{listing}
\begin{listing}[tbp]
\begin{code}
  data AD a = AD a a deriving (Show, Eq, Ord)
  instance Floating a => Floating (AD a) where
    exp (AD f f') = AD (exp f) (f' * exp f)
    sin (AD f f') = AD (sin f) (f' * cos f)
    log (AD f f') = AD (log f) (f' / f)
    ...
\end{code}
\caption{The definition of \texttt{AD}\label{lst:def-AD}}
\end{listing}

For example, a simple forward-mode AD implemented as a dual number can be defined as in \Cref{lst:def-AD}.
The data-type \hask{AD} encapsulates a value of some univariate function and its first-order differential coefficient and calculates the result using the Chain Rule, using \hask{Floating}-operations on the coefficient \hask{a}.

So our goal is to implement \hask{STower n a} data-type conveying information of \emph{all} the higher-order derivatives of an $n$-variate smooth function on \hask{a}, which has \hask{Floating (STower n a)} instances for all \hask{Floating a} and $n$.
In addition, we demand the implementation to be \emph{succinct} and \emph{efficient}, in a sense that it avoids equivalent calculation as much as possible.
For example, if we want to calculate $f_{xy^2}(a,b)$ and $f_{x^2y}(a,b)$ for some smooth $f: \R^2 \to \R$, it should calculate the derivatives up to $f_{xy}$ at most once and share their results in computing both $f_{xy^2}$ and $f_{x^2y}$; in other words, results up to $f_{xy}$ must be \emph{memoised}.

\begin{figure}[tbp]
  \centering
  \begin{tikzpicture}[
    level/.style={level distance=1cm},
    level 1/.style={sibling distance=3.75cm},
    level 2/.style={sibling distance=1.2cm},
    level 3/.style={sibling distance=4mm}
    ]
    \newcommand{\ba}{\boldsymbol{a}}
    \node (fa) {$f(\ba)$}
      child{ 
        node (fxa) {$f_x(\ba)$}
          child { 
            node (fxxa) {$f_{xx}(\ba)$}
            child {node{} edge from parent[thick,dotted]}
            child {node{} edge from parent[thick,dotted]}
            child {node{} edge from parent[thick,dotted]}
          }
          child {
            node (fxya) {$f_{xy}(\ba)$}
            child {node{} edge from parent[thick,dotted]}
            child {node{} edge from parent[thick,dotted]} 
            child {node{} edge from parent[thick,dotted]}
          }
          child { 
            node (fxza) {$f_{xz}(\ba)$}
            child {node{} edge from parent[thick,dotted]}
            child {node{} edge from parent[thick,dotted]}
            child {node{} edge from parent[thick,dotted]}
          }
      }
      child {
        node (fya) {$f_y(\ba)$}
        child { 
          node {$f_{yx}(\ba)$} 
          child {node{} edge from parent[thick,dotted]}
          child {node{} edge from parent[thick,dotted]}
          child {node{} edge from parent[thick,dotted]}
        }
        child { 
          node {$f_{yy}(\ba)$}
          child {node{} edge from parent[thick,dotted]}
          child {node{} edge from parent[thick,dotted]}
          child {node{} edge from parent[thick,dotted]}
        }
        child { 
          node {$f_{yz}(\ba)$} 
          child {node{} edge from parent[thick,dotted]}
          child {node{} edge from parent[thick,dotted]}
          child {node{} edge from parent[thick,dotted]}
        }
      }
      child {
        node (fza) {$f_z(\ba)$}
        child {
            node {$f_{zx}(\ba)$}
            child {node{} edge from parent[thick,dotted]}
            child {node{} edge from parent[thick,dotted]}
            child {node{} edge from parent[thick,dotted]}
        }
        child {
          node {$f_{zy}(\ba)$}
          child {node{} edge from parent[thick,dotted]}
          child {node{} edge from parent[thick,dotted]}
          child {node{} edge from parent[thick,dotted]}
        }
        child {
            node {$f_{zz}(\ba)$}
            child {node{} edge from parent[thick,dotted]}
            child {node{} edge from parent[thick,dotted]}
            child {node{} edge from parent[thick,dotted]}
        }
      };
  \end{tikzpicture}
  \caption{Trivariate case, first trial\label{fig:tree-simpl}}
\end{figure}
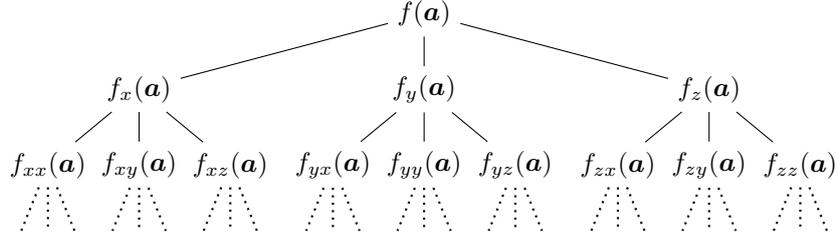

To that end, we employ an infinite tree representation.
The main idea is to use an $n$-ary infinite tree to express a (piecewise) smooth functions: the root node corresponds to the value $f(\boldsymbol{a})$, and its child in the $n$\textsuperscript{th} branch corresponds to $\partial x_n f(\boldsymbol{a})$.
The intuition in the trivariate case is depected in \Cref{fig:tree-simpl}.
This can be viewed as a nested trie (or prefix-tree) for memoising functions with $n$-many natural number arguments.
Actually, this representation is isomorphic to the one we can obtain when \hask{Sparse} tower from \texttt{ad} package~\cite{Kmett:2010aa} to the fixed-length vectors.
However, as we assume $f$ to be (piecewise) smooth, there are space for optimisation.
That is, in the above representation, $f_{xy}$ and $f_{yx}$ must almost always coincide except on non-smooth points.
In other words, we can assume partial differential operators to be almost always commutative on inputs.
In many applications, the value on the non-smooth points is negligible and hence we must use more \emph{succinct} representation making use of the commutativity.

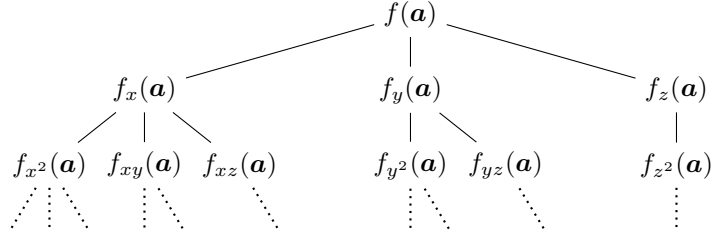
\begin{figure}[tbp]
  \centering
  \begin{tikzpicture}[
    level/.style={level distance=1cm},
    level 1/.style={sibling distance=3.5cm},
    level 2/.style={sibling distance=1.25cm},
    level 3/.style={sibling distance=6mm}
    ]
    \newcommand{\ba}{\boldsymbol{a}}
    \node (fa) {$f(\ba)$}
      child{ 
        node (fxa) {$f_x(\ba)$}
          child { 
            node (fxxa) {$f_{x^2}(\ba)$}
            child {node{} edge from parent[thick,dotted]}
            child {node{} edge from parent[thick,dotted]}
            child {node{} edge from parent[thick,dotted]}
          }
          child {
            node (fxya) {$f_{xy}(\ba)$}
            child[missing]
            child {node{} edge from parent[thick,dotted]}
            child {node{} edge from parent[thick,dotted]}
          }
          child { 
            node (fxza) {$f_{xz}(\ba)$}
            child[missing]
            child[missing]
            child { node{} edge from parent[thick,dotted] }
          }
      }
      child {
        node (fya) {$f_y(\ba)$}
        child[missing]
        child { node {$f_{y^2}(\ba)$} child[missing] 
          child {node{} edge from parent[thick,dotted]}
          child { node{} edge from parent[thick,dotted]}
        }
        child { node {$f_{yz}(\ba)$} child[missing] child[missing]
          child {node{} edge from parent[thick,dotted]} }
      }
      child {
        node (fza) {$f_z(\ba)$}
        child {
            node {$f_{z^2}(\ba)$}
            child { node{} edge from parent[thick,dotted] }
        }
      };
  \end{tikzpicture}
  \caption{Trivariate case, succinct version\label{fig:tree}}
\end{figure}

The idea is simple: if once one goes down $i$\textsuperscript{th} path, we can only choose $j$\textsuperscript{th} branches for $j \geq i$.
This trick is illustrated in \Cref{fig:tree} for trivariate case.
This can be seen as a special kind of an infinite trie (or prefix-tree) of alphabets $\partial_{x_i}$ with available letter eventually decreasing.

\begin{figure}[tbp]
  \centering
  \begin{tikzpicture}[
    level/.style={level distance=1cm},
    level 1/.style={sibling distance=8cm},
    level 2/.style={sibling distance=4.5cm},
    level 3/.style={sibling distance=2cm},
    level 4/.style={sibling distance=1.25cm},
    level 5/.style={sibling distance=0.75cm}
    ]
    \newcommand{\ba}{\boldsymbol{a}}
    \node (fa) {$f(\ba)$}
      child{ 
        node (fxa) {$f_x(\ba)$}
          child { 
            node (fxxa) {$f_{x^2}(\ba)$}
            child {node{} edge from parent[thick,dotted]}
            child {
              node{$f_{x^2}(\ba)$}
              child {
                node {$f_{x^2y}(\ba)$}
                child {node{} edge from parent[thick,dotted]}
                child {node{} edge from parent[thick,dotted]}
              }
              child { node { $f_{x^2}(\ba)$ }
                child { 
                  node { $f_{x^2z}(\ba)$ } 
                  child {node{} edge from parent[thick,dotted]}
                }
              }
            }
          }
          child {
            node {$f_x(\ba)$}
            child {
              node (fxya) {$f_{xy}(\ba)$}
              child {
                node{$f_{xy^2}$}
                child { node{} edge from parent[thick,dotted] }
                child { node{} edge from parent[thick,dotted] }
              }
              child {
                node {$f_{xy}(\ba)$}
                child { 
                  node {$f_{xyz}$}
                  child {node{} edge from parent[thick,dotted]}
                }
              }
            }
            child { 
              node (fxza) {$f_{x}(\ba)$}
              child{
                node {$f_{xz}(\ba)$}
                child {node{} edge from parent[thick,dotted]}
              }
            }
          }
      }
      child
      {
        node {$f(\ba)$}
        child {
          node (fya) {$f_y(\ba)$}
          child { node {$f_{y^2}(\ba)$} 
            child {node{} edge from parent[thick,dotted]}
            child {node{} edge from parent[thick,dotted]}
            }
          child { node {$f_{y}(\ba)$} 
            child {
              node {$f_{yz}(\ba)$}
              child{node{} edge from parent[thick,dotted]}
            }
          }
        }
        child {
          node (fza) {$f(\ba)$}
          child {
              node {$f_{z}(\ba)$}
              child { node {$f_{z^2}(\ba)$}
                child{node{} edge from parent[thick,dotted]}
              }
          }
        }
      }
      ;
  \end{tikzpicture}
  \caption{Trivariate case, tweaked succinct version\label{fig:tree-tweaked}}
\end{figure}
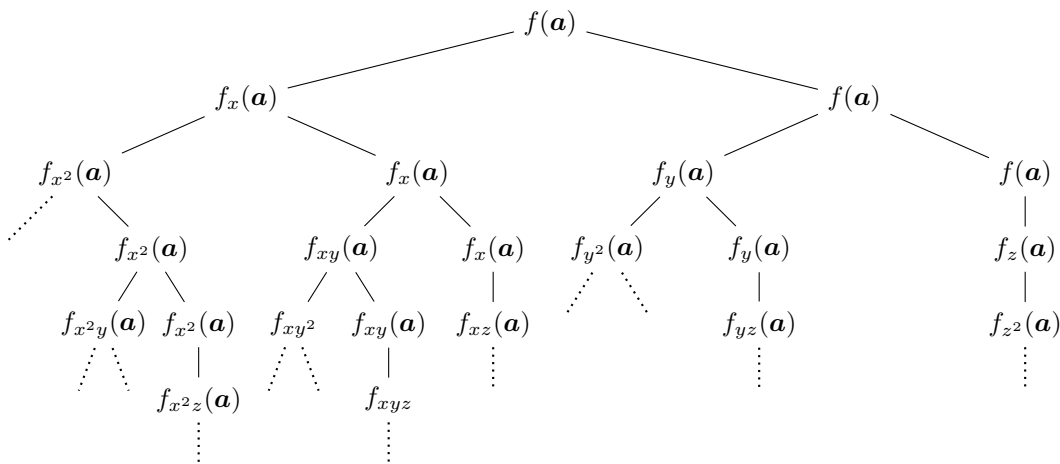

We further tweak this representation to make data-type definition and algorithm simple (\Cref{fig:tree-tweaked}).
This has two advantages:

\begin{enumerate}
\item We can directly express the tree structure purely as an algebraic data-type (ADT); we need to use fixed-length vectors here otherwise.\label{item:adt-friendly}
\item Derivatives can be computed simply by induction on the number of variables.\label{item:recurse}
\end{enumerate}

To see these advantages, we now turn to the Haskell implementation (\Cref{lst:ops-def}).
\begin{listing}[tb]
\begin{code}
data STower n a where
  ZS :: !a -> STower 0 a
  SS :: !a -> STower (n + 1) a -> STower n a -> STower (n + 1) a

instance Num a => Num (STower n a) where
  ZS a + ZS b = SZ (a + b)
  SS f df dus + SS g dg dvs = SS (f + g) (df + dg) (dus + dvs)

  ZS a * ZS b = ZS (a * b)
  SS f df dus * SS g dg dvs = (f * g) (f * dg + df * g) (dus * dvs)
  ...

instance Fractional a => Fractional (STower n a) where
  ZS a / ZS b = ZS (a / b)
  SS f df dus / SS g dg dvs = SS (f / g) (df / g - f * dg / g^2) (dus / dvs)

instance Floating a => Floating (STower n a) where
  log (ZS a) = ZS (log a)
  log (SS f df dus) = SS (log f) (df / f) (log dus)
  sin (ZS a) = ZS (sin a)
  sin (SS f df dus) = SS (sin f) (df * cos f) (sin dus)
  cos (ZS a) = ZS (cos a)
  cos (SS f df dus) = SS (cos f) (- df * sin f) (cos dus)
  exp (ZS a) = ZS (exp a)
  exp (SS f df dus) = SS (exp f) (df * exp f) (exp dus)
  ...
\end{code}
\caption{Definitions of operations of \texttt{STower}\label{lst:ops-def}}
\end{listing}
\hask{STower n a} is the type corresponding to the $n$-variate formal power series ring.
Instance declarations implements functions and their derivatives on \hask{STower n a}.
It might seems circular at the first glance, but these definitions works several reasons.
For example, the definitions of $\sin$ and $\cos$ works because:
\begin{enumerate}
\item The calls of $\sin$ and $\cos$ in the first arguments (\hask{sin f} and \hask{cos f}) are those on the value \hask{f} in the coefficient field \hask{a}, not \hask{STower n a}, whose implementation is already given.
\label{itm:call-coeff}
\item The second calls of $\sin$ and $\cos$ (\hask{df * cos f} and \hask{-df * sin x}) is those on the \hask{STower n a} itself and results in an infinite loop.
However, we are constructing an infinite trees that carries \emph{all} the partial derivative coefficients, and those coefficients are stored as the first argument of \hask{SS}, which has definite values by the discussion in \Cref{itm:call-coeff}.
\label{itm:call-deriv}
\item The direct calls in final arguments (\hask{sin dus} and \hask{cos dus}) seems looping, but they are indeed on power series with strictly less variable,that is, on \hask{STower (n - 1) a}, instead of \hask{STower n a}.
As these arity strictly decreases, this stops after finite steps.
\label{itm:rec-decr}
\end{enumerate}
Note that \Cref{itm:call-deriv} works also because both \hask{sin} and \hask{cos} are members of the \hask{Floating} class.
In other words, member functions of the \hask{Floating} class is closed under derivatives.
More generally, if the class \hask{c a} provides a family of numerical functions on \hask{c} which is closed under derivatives (together with functions from their superclasses), we can likewise derive the instance \hask{c (STower n a)} as above.
This idea is embodied in \hask{liftSTower} function in \Cref{lst:lift}, and the usage is illustrated by the alternative definitions of instances that follows.

\begin{listing}[tbp]
\begin{code}
liftSTower
  :: forall c n a. (KnownNat n, c a, forall x k. c x => c (STower k x) )
  => (forall x. c x => x -> x)
      -- ^ Function
  -> (forall x. c x => x -> x)
      -- ^ its first-order derivative
  -> STower n a
  -> STower n a
liftSTower f df (ZS a) = ZS (f a)
liftSTower f df x@(SS a da dus) = SS (f a) (da * df x) (f df dus)

instance Floating a => Floating (STower n a) where
  log = liftSTower @Floating log recip
  sin = liftSTower @Floating sin cos
  cos = liftSTower @Floating cos (negate . sin)
  exp = liftSTower @Floating exp exp
\end{code}
\caption{Helper function for implementing derivatives\label{lst:lift}}
\end{listing}

One might note that \Cref{itm:rec-decr} and the last argument in the \hask{SS}-clause in \hask{liftSmooth} is really simple recursion.
We adopted the tweaked representation as depicted in \Cref{fig:tree-tweaked}, instead of \Cref{fig:tree}, for this simplicity.
If we make branching $n$-ary as in \Cref{fig:tree}, the implementation gets more complicated as the number of variable decreases.
One might worry that the same coefficient gets duplicated among multiple branches, but if we choose a language with a sharing (like Haskell), these value are shared across the siblings\footnote{We could achieved this also in more low-level langauges, such as C/C++, by representing each node as a pointer instead of a direct value.}.

\section{Benchmarks}
\label{sec:bench}
We report several benchmarks on the proposed method with the existing implementation in the \texttt{ad}~\cite{Kmett:2010aa} package.

The code used in the benchmarks is implemented as a Haskell library and hosted on GitHub\footnote{\url{https://github.com/konn/smooth/tree/d3386a15c97f23d4071b09f97b7bc87f2c4b1da4}}.
All the benchmark code was compiled with GHC 8.10.4 with the flag \texttt{-threaded -O2}.
We ran the benchmark suites on a virtual Linux environment available on GitHub Actions (Standard\_DS2\_v2 Azure instance) with two Intel Xeon Platinum 8171M virtual CPUs (2.60GHz) and 7 GiB of RAM.
The Gauge framework~\cite{Hanquez:2019wk} was used to report the run-time speed.

\subsection{Tower Automatic Differentiation}
In this subsection, we compare the run-time speed and memory consumption of the existing multivariate tower AD and the proposed method.
In univariate case, we compare our proposed method with two existing implementations: \hask{Sparse} and \hask{diffs} both from \texttt{ad} package.
The former is the generic implementation of multivariate tower AD, and the former is speciailised implementation for the univariate case.

\begin{figure}[htbp]
\begin{center}
    \begin{tikzpicture}
    \begin{axis}[%
        legend columns=-1,
        legend entries={{\texttt{Sparse}},\textt{diffs},our method},
        legend to name=leg:time-tower,
        xlabel={Order},ylabel={Time (sec)},%
        title=Identity,%
      ]
      \addplot[mark=*] table[col sep=comma,x=size,y=AD] 
        {bench-results/multdiffupto-speed/identity.csv};
      \addplot[mark=o] table[col sep=comma,x=size,y=AD-diffs] 
        {bench-results/multdiffupto-speed/identity.csv};
      \addplot[mark=x] table[col sep=comma,x=size,y=STower] 
        {bench-results/multdiffupto-speed/identity.csv};
    \end{axis}
  \end{tikzpicture}
  \begin{tikzpicture}
  \begin{axis}[%
      xlabel={Order},ylabel={Time (sec)},%
      title=$e^x$,%
    ]
    \addplot[mark=*] table[col sep=comma,x=size,y=AD] 
      {bench-results/multdiffupto-speed/exp-x.csv};
    \addplot[mark=o] table[col sep=comma,x=size,y=AD-diffs] 
      {bench-results/multdiffupto-speed/exp-x.csv};
    \addplot[mark=x] table[col sep=comma,x=size,y=STower] 
      {bench-results/multdiffupto-speed/exp-x.csv};
  \end{axis}
\end{tikzpicture}

\begin{tikzpicture}
  \begin{axis}[%
      xlabel style={align=center},
      xlabel={Total degrees:\\$(0, 1), (2, 0), (2, 1), (2, 2), (3, 2)$\\
        $(4, 2), (3, 4), (5, 3), (3, 6), (6, 4)$},ylabel={Time (sec)},%
      title=$\sin x e^{y^2}$,%
    ]
    \addplot[mark=*] table[col sep=comma,x=size,y=AD] 
      {bench-results/multdiffupto-speed/sin-x-exp-y2.csv};
    \addplot[mark=x] table[col sep=comma,x=size,y=STower] 
      {bench-results/multdiffupto-speed/sin-x-exp-y2.csv};
  \end{axis}
\end{tikzpicture}
\begin{tikzpicture}
  \begin{axis}[%
      xlabel style={align=center},
      xlabel={Total degrees:\\
      {$\scriptsize (0, 0, 1), (1, 0, 1), (0, 1, 2), (1, 2, 1), (0, 3, 2), (2, 2, 2)$}\\
      {$\scriptsize (3, 2, 2), (3, 4, 1), (5, 3, 1), (2, 3, 5), (5, 4, 2), (3, 4, 5)$}},%
      ylabel={Time (sec)},%
      title=$\sin x e^{y^2 + z}$,%
    ]
    \addplot[mark=*] table[col sep=comma,x=size,y=AD] 
      {bench-results/multdiffupto-speed/sin-x-exp-y2-z.csv};
    \addplot[mark=x] table[col sep=comma,x=size,y=STower] 
      {bench-results/multdiffupto-speed/sin-x-exp-y2-z.csv};
  \end{axis}
\end{tikzpicture}

\ref*{leg:time-tower}
\end{center}

\caption{Speed benchmarks. \texttt{Sparse} and \texttt{diffs} are existing implementation in \texttt{ad} package.\label{fig:tower-time-bench}}
\end{figure}
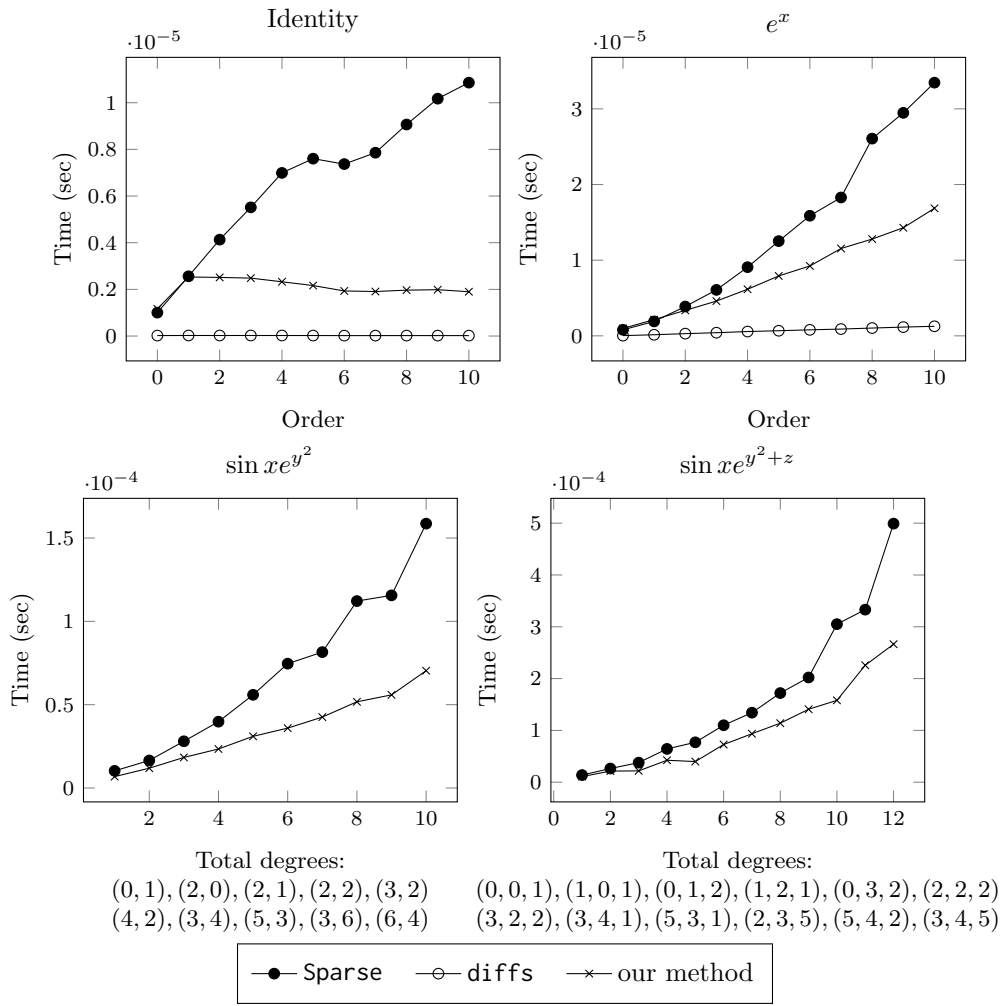

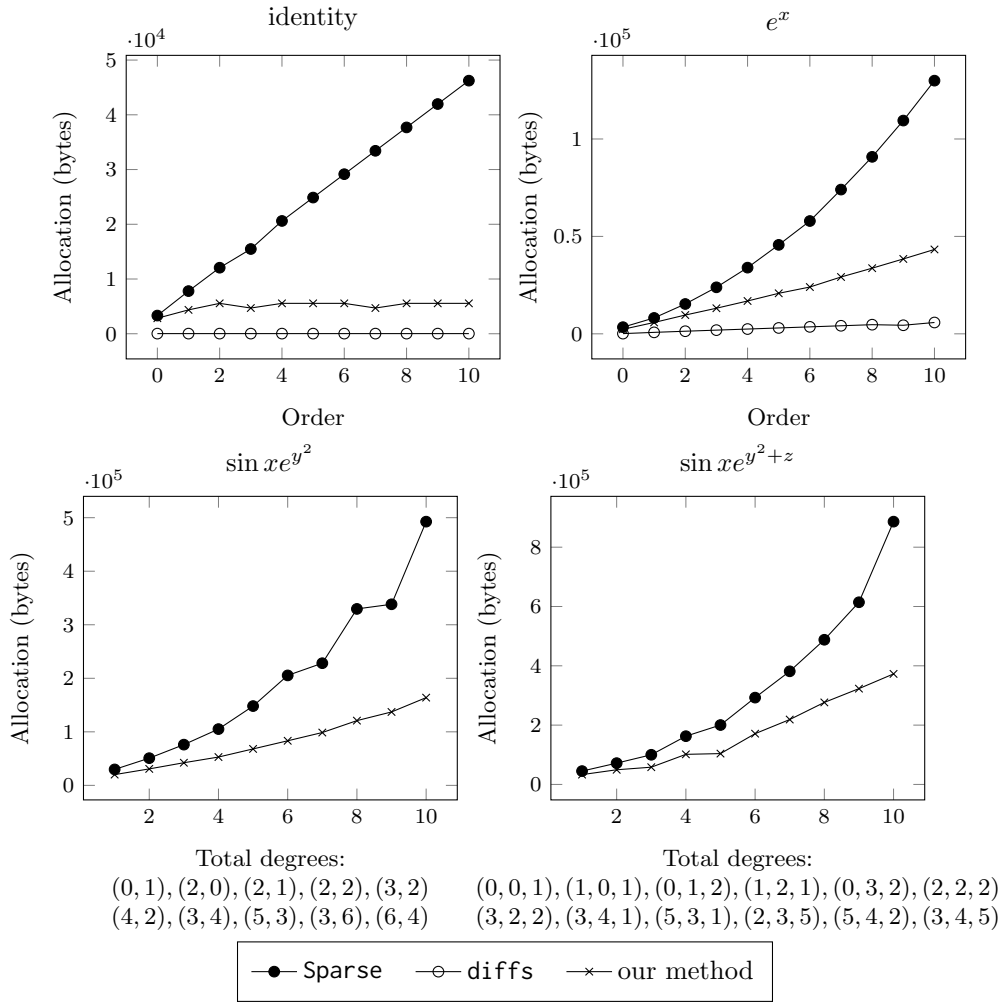
\begin{figure}[htbp]
  \begin{center}
      \begin{tikzpicture}
      \begin{axis}[%
          legend columns=-1,
          legend entries={{\texttt{Sparse}},\textt{diffs},our method},
          legend to name=leg:heap-tower,%
          xlabel={Order},ylabel={Allocation (bytes)},%
          title=identity,%
        ]
        \addplot[mark=*] table[col sep=comma,x=size,y=AD] 
          {bench-results/multdiffupto-heap/identity.csv};
        \addplot[mark=o] table[col sep=comma,x=size,y=AD-diffs] 
          {bench-results/multdiffupto-heap/identity.csv};
        \addplot[mark=x] table[col sep=comma,x=size,y=STower] 
          {bench-results/multdiffupto-heap/identity.csv};
      \end{axis}
    \end{tikzpicture}
    \begin{tikzpicture}
    \begin{axis}[%
        xlabel={Order},ylabel={Allocation (bytes)},%
        title=$e^x$,%
      ]
      \addplot[mark=*] table[col sep=comma,x=size,y=AD] 
        {bench-results/multdiffupto-heap/exp-x.csv};
      \addplot[mark=o] table[col sep=comma,x=size,y=AD-diffs] 
        {bench-results/multdiffupto-heap/exp-x.csv};
      \addplot[mark=x] table[col sep=comma,x=size,y=STower] 
        {bench-results/multdiffupto-heap/exp-x.csv};
    \end{axis}
  \end{tikzpicture}
  
  \begin{tikzpicture}
    \begin{axis}[%
        xlabel style={align=center},
        xlabel={Total degrees:\\$(0, 1), (2, 0), (2, 1), (2, 2), (3, 2)$\\
          $(4, 2), (3, 4), (5, 3), (3, 6), (6, 4)$},ylabel={Allocation (bytes)},%
        title=$\sin x e^{y^2}$,%
      ]
      \addplot[mark=*] table[col sep=comma,x=size,y=AD] 
        {bench-results/multdiffupto-heap/sin-x-exp-y2.csv};
      \addplot[mark=x] table[col sep=comma,x=size,y=STower] 
        {bench-results/multdiffupto-heap/sin-x-exp-y2.csv};
    \end{axis}
  \end{tikzpicture}
  \begin{tikzpicture}
    \begin{axis}[%
        xlabel style={align=center},
        xlabel={Total degrees:\\
        {$\scriptsize (0, 0, 1), (1, 0, 1), (0, 1, 2), (1, 2, 1), (0, 3, 2), (2, 2, 2)$}\\
        {$\scriptsize (3, 2, 2), (3, 4, 1), (5, 3, 1), (2, 3, 5), (5, 4, 2), (3, 4, 5)$}},%
        ylabel={Allocation (bytes)},%
        title=$\sin x e^{y^2 + z}$,%
      ]
      \addplot[mark=*] table[col sep=comma,x=size,y=AD] 
        {bench-results/multdiffupto-heap/sin-x-exp-y2-z.csv};
      \addplot[mark=x] table[col sep=comma,x=size,y=STower] 
        {bench-results/multdiffupto-heap/sin-x-exp-y2-z.csv};
    \end{axis}
  \end{tikzpicture}

  \ref*{leg:heap-tower}
\end{center}
  \caption{Heap benchmarks. \texttt{Sparse} and \texttt{diffs} are existing implementation in \texttt{ad} package.\label{fig:tower-heap-bench}}
\end{figure}

\Cref{fig:tower-time-bench,fig:tower-heap-bench} show the run-time speed and heap allocation of the existing implementations (\hask{Sparse} and \hask{diffs} only for univariate case), and the proposed method.
In univaricate case, the existing implementation for univariate tower AD provided, i.e.\ \hask{diffs} from \texttt{ad} package, peforms the best both in terms of run-time speed and memory allocation.
Notably, our implementation outperforms \hask{Sparse} both in time and space.
In particular, although \hask{Sparse} performs linearly both in time and space when applied to the identity, our method performs eventually constantly.
This is because our actual implementation employs some heuristics to detect a function whose higher derivatives are eventually zero.

Compared to \hask{Sparse}, our proposed method performs always better both in univariate and multivariate cases.
In particular, \hask{Sparse} presents a steep quadratic growth both in time and space, our proposed method peroforms almost linearly.

In summary, our method performs really well both in terms of time and space in the multivariate case, although we need more optimisation in univariate cases.

\subsection{Weil Algebra Computation}
In this subsection, we compare the performances of Tower AD applied to Weil algebra computation as described in \cite{Ishii:2021vw}.

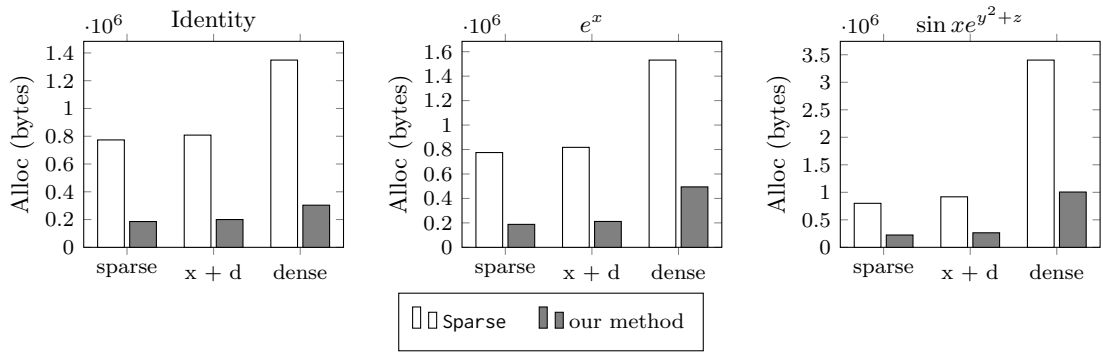
\begin{figure}[btp]
  \begin{center}
    \begin{tikzpicture}
      \begin{axis}[%
          footnotesize,
          legend columns=-1,
          legend entries={{\texttt{Sparse}},our method},
          legend to name=leg:time-weil,
          ylabel={Time (sec)},%
          title=Identity,%
          symbolic x coords={sparse,x + d,dense},
          enlarge x limits=0.25,
          xtick=data,
          ymin=0,
          ybar,
        ]
        \addplot[black,fill=white] table[col sep=comma,x=size,y=AD] 
          {bench-results/liftWeil-speed/identity.csv};
        \addplot[black,fill=gray] table[col sep=comma,x=size,y=STower] 
          {bench-results/liftWeil-speed/identity.csv};
      \end{axis}
    \end{tikzpicture}%
    \quad
    \begin{tikzpicture}
      \begin{axis}[%
          footnotesize,
          ylabel={Time (sec)},%
          title={$e^x$},%
          symbolic x coords={sparse,x + d,dense},
          enlarge x limits=0.25,
          xtick=data,
          ymin=0,
          ybar,
        ]
        \addplot[black,fill=white] table[col sep=comma,x=size,y=AD] 
          {bench-results/liftWeil-speed/exp-x.csv};
        \addplot[black,fill=gray] table[col sep=comma,x=size,y=STower] 
          {bench-results/liftWeil-speed/exp-x.csv};
      \end{axis}
    \end{tikzpicture}
    \quad%
    \begin{tikzpicture}
      \begin{axis}[%
          footnotesize,
          ylabel={Time (sec)},%
          title={$\sin x e^{y^2 + z}$},%
          symbolic x coords={sparse,x + d,dense},
          enlarge x limits=0.25,
          xtick=data,
          ymin=0,
          ybar,
        ]
        \addplot[black,fill=white] table[col sep=comma,x=size,y=AD] 
          {bench-results/liftWeil-speed/sin-x-exp-y2-z.csv};
        \addplot[black,fill=gray] table[col sep=comma,x=size,y=STower] 
          {bench-results/liftWeil-speed/sin-x-exp-y2-z.csv};
      \end{axis}
    \end{tikzpicture}
  
  \ref*{leg:time-weil}

  \texttt{Sparse} is the existing implementation in \texttt{ad} package.
  As an input, the ``sparse'' gives $1$ , ``x + d'' as it is, and ``dense'' $x + (\text{sum of all nonzero basis})$.
  
  \caption{Speed benchmarks for Weil algebra $W = \R[x,y]/(x^3 - y^2, y^3)$. \label{fig:weil-time-bench}}
  \end{center}
\end{figure}

\begin{figure}[btp]
  \begin{center}
    \begin{tikzpicture}
      \begin{axis}[%
          footnotesize,
          legend columns=-1,
          legend entries={{\texttt{Sparse}},our method},
          legend to name=leg:time-weil,
          ylabel={Alloc (bytes)},%
          title=Identity,%
          symbolic x coords={sparse,x + d,dense},
          enlarge x limits=0.25,
          xtick=data,
          ymin=0,
          ybar,
        ]
        \addplot[black,fill=white] table[col sep=comma,x=size,y=AD] 
          {bench-results/liftWeil-heap/identity.csv};
        \addplot[black,fill=gray] table[col sep=comma,x=size,y=STower] 
          {bench-results/liftWeil-heap/identity.csv};
      \end{axis}
    \end{tikzpicture}
    \quad
    \begin{tikzpicture}
      \begin{axis}[%
          footnotesize,
          ylabel={Alloc (bytes)},%
          title={$e^x$},%
          symbolic x coords={sparse,x + d,dense},
          enlarge x limits=0.25,
          xtick=data,
          ymin=0,
          ybar,
        ]
        \addplot[black,fill=white] table[col sep=comma,x=size,y=AD] 
          {bench-results/liftWeil-heap/exp-x.csv};
        \addplot[black,fill=gray] table[col sep=comma,x=size,y=STower] 
          {bench-results/liftWeil-heap/exp-x.csv};
      \end{axis}
    \end{tikzpicture}
    \quad%
    \begin{tikzpicture}
      \begin{axis}[%
          footnotesize,
          ylabel={Alloc (bytes)},%
          title={$\sin x e^{y^2 + z}$},%
          symbolic x coords={sparse,x + d,dense},
          enlarge x limits=0.25,
          xtick=data,
          ymin=0,
          ybar,
        ]
        \addplot[black,fill=white] table[col sep=comma,x=size,y=AD] 
          {bench-results/liftWeil-heap/sin-x-exp-y2-z.csv};
        \addplot[black,fill=gray] table[col sep=comma,x=size,y=STower] 
          {bench-results/liftWeil-heap/sin-x-exp-y2-z.csv};
      \end{axis}
    \end{tikzpicture}
  
  \ref*{leg:time-weil}

  \texttt{Sparse} is the existing implementation in \texttt{ad} package.
  As an input, the ``sparse'' gives $1$ , ``x + d'' as it is, and ``dense'' $x + (\text{sum of all nonzero basis})$.
  
  \caption{Heap benchmarks for Weil algebra $W = \R[x,y]/(x^3 - y^2, y^3)$. \label{fig:weil-heap-bench}}
  \end{center}
\end{figure}

\Cref{fig:weil-time-bench,fig:weil-heap-bench} present the time and space performance of Weil algebra computation based on the existing implementation (\texttt{Sparse}) and the proposed method.
In particular, we evaluated three functions (the identity, $e^x$, and $\sin x \cdot e^{y^2 + z}$) on a Weil algebra $W = \R[x,y]/(x^3 - y^2, y^3)$.
We feed three types of inputs: a sparse input ($1$), $1 + d$, and dense input $1 + d_1 + d_2 + d_1^2 + d_1 d_2 + d_2^2$.
In any case, our proposed method largely outperforms the existing implementation.

\section{Conclusion}
\label{sec:concl}
We presented a succinct and efficient implementation of a multivariate lazy forward-mode tower automatic differentiation.
This can be viewed as a mixture of existing methods but optimised exploiting the commutativity of partial differential operators.
The basic idea is to store all the partial derivatives in some kind of a prefix-tree, in which the number of branching will eventually decrease as the right branch is chosen.
We applied laziness and the advanced type-system in Haskell.

Our implementation performs particularly well in multivariate cases and gives pleasing improvements both in time and space when applied to Weil algebra computation as described in \cite{Ishii:2021vw}.
For univariate case, however, there is much room for improvements.
It might be a good future work to explore the special treatment in the univariate case to remove overheads.

\begin{acknowledgements}
  The author would like to thank Prof.\ Akira Terui for encouraging me to participate in the RIMS workshop.
\end{acknowledgements}

\printbibliography

\end{document}